\documentclass[12pt]{iopart}

\usepackage{graphicx}
\usepackage{amsmath}
\usepackage{amssymb}
\usepackage{color}
\usepackage{caption}
\usepackage{subcaption}
\newcommand{\red}[1]{\textcolor{black}{#1}}

\begin{document}

\title[On neoclassical impurity transport in stellarator geometry]{On neoclassical 
impurity transport in stellarator geometry}

\author{J. M. Garc\'ia-Rega\~na, R. Kleiber, C. D. Beidler, Y. Turkin, H. Maa\ss berg and  P. Helander }

\address{Max-Planck-Institut f\"ur Plasmaphysik, 
EURATOM-Assoziation, Wendelsteinstr. 1, 17491 Greifswald, Germany}

\ead{jose.regana@ipp.mpg.de}

\begin{abstract}
The impurity dynamics in stellarators has become an issue of 
moderate concern due to the
inherent tendency of the impurities to accumulate 
in the core when the neoclassical 
ambipolar radial electric field points
radially inwards (ion root regime). This accumulation can lead to  
collapse of the plasma due to radiative losses, and thus limit
high performance plasma discharges in non-axisymmetric devices.\\
A quantitative description 
of the neoclassical impurity transport is complicated by the breakdown of the
assumption of small $\mathbf{E}\times \mathbf{B}$ drift and trapping due to the electrostatic
potential variation on a flux surface $\tilde{\Phi}$ compared 
to those due to the magnetic field gradient.
The present work examines the impact of this potential variation on neoclassical 
impurity transport in the Large Helical Device (LHD) stellarator.
It shows that the neoclassical impurity transport can be strongly affected 
by $\tilde{\Phi}$.
The central numerical tool used 
is the $\delta f$ particle in cell (PIC) Monte Carlo code \texttt{EUTERPE}.
The $\tilde{\Phi}$ used in the calculations is provided by
the neoclassical code \texttt{GSRAKE}. The possibility of obtaining
a more general $\tilde{\Phi}$ self-consistently with \texttt{EUTERPE}
is also addressed and a preliminary calculation is presented.
\end{abstract}

\section{Introduction}

Thermonuclear fusion would benefit from the 
achievement of quasi steady-state magnetic plasmas confinement
with similar characteristics to those
expected in a future reactor. In this respect the stellarator
concept has an advantage over the pulsed tokamak.
On the other hand the neoclassical transport exhibited by the former 
at low collisionality in the absence of electric fields, in the so-called
$1/\nu$ regime, is considerably larger. The reason for this
unfavourable behaviour is that in contrast to the axisymmetric case, where
the collisionless trajectories of the trapped particle are confined, 
the orbits of particles trapped in the helical magnetic wells are generally not
confined. 
This situation leads to a different radial transport rate 
for each species in the plasma and necessitates 
a radial electric field $\mathbf{E}_{r}=-\nabla \Phi_{0}$ that restores ambipolarity. Here
$\Phi_{0}=\Phi_{0}(s)$ is the lowest order electrostatic potential and
only depends on the flux surface label. In the present work 
we use $s=\psi/\psi_{0}$, with $\psi$ is the toroidal flux and
$\psi_{0}$ the toroidal flux at the last closed magnetic surface. 
This label can be written in terms of the effective radius $r$ as
$s\approx(r/a)^2$ with $a$ the minor radius of the plasma.
In the present context
the basic ordering parameter, $\delta\equiv\rho/L$, is the Larmor radius $\rho$
normalized to a typical macroscopic variation length scale $L$. 
Consequently
the distribution function is expressible up to first order as
$f\approx f_{0}+\delta f$, with $\delta f/f_{0}\sim O(\delta)$.
In standard conditions \red{$\mathbf{E}_{r}$}
points inwards (ion root regime) and predicts accumulation of impurities 
\cite{Maassberg_ppcf_41_1999}. The subsequent increase of radiative
losses from the core can cause the collapse of the plasma 
\cite{Hirsch_ppcf_50_053001_2008, Nakamura_ppcf_44_2121_2002},
and, in the worst case, endanger the capability 
of the stellarator to confine it in steady-state.\\
A quantitative and comprehensive description 
of the impurity dynamics requires not only
the ambipolar radial electric field but also 
the electrostatic potential $\tilde{\Phi}=\Phi - \Phi_{0}$ determined by $\delta f$.
This portion of the total electrostatic potential $\Phi$, whose 
explicit solution has been obtained in the past \cite{Dobrott_pp_11_211_1969, Mynick_pf_27.8_1984} 
and pointed out to have a modest impact 
on the transport of the bulk species 
\cite{Mynick_pf_27.8_1984, Ho_pf_30.2_1987, Beidler_isw_2005} 
has been traditionally neglected.
This rests on two assumptions.
First, the radial $\mathbf{E}\times \mathbf{B}$ drift arising from 
$\tilde{\Phi}$,

\begin{equation}
  \mathbf{v}_{\tilde{\Phi}}=-\frac{\nabla\tilde{\Phi}\times\mathbf{b}}{B},
\end{equation}
is assumed to be smaller than the curvature and \textit{grad-B} drifts.
In the vacuum approximation these are given by,

\begin{equation}
  \mathbf{v}_{\mathrm{d}}=\frac{m}{q}\frac{\mu B + v_{\|}^2}{B^2}\mathbf{b}\times\nabla B,
\end{equation}
with $m$ the mass of the particle, $q=Ze$ its electric charge, 
$Z$ is the charge number, $e$ is the unit charge absolute value, $B$ is the 
magnetic field strength, $\mathbf{b}$ is a unitary vector pointing in the direction of the 
magnetic field line, $\mu=v_{\bot}^{2}/2B$ is the magnetic moment, 
$v_{\|}$ is the parallel velocity and $v_{\bot}$ is the perpendicular one.
It can be shown that the ratio between the absolute values of $\mathbf{v}_{\tilde{\Phi}}$ 
and $\mathbf{v}_{\mathrm{d}}$ is thus of order

\begin{equation}
  \label{drift_limit}
  \frac{v_{\tilde{\Phi}}}{v_{\mathrm{d}}}\sim\frac{Z e\tilde{\Phi}}{T}\frac{R}{a},
\end{equation}
where $R$ is the major radius of the device, and
it is assumed that the typical
variation length scales for $B$ and $\tilde{\Phi}$ are similar to $R$ and $a$,
respectively. This ratio can be acceptably small at low values of
$Z$, but is usually considerable for heavy impurities.\\
Second, the parallel acceleration $a_{\|}=-(q/m)\mathbf{b}\cdot\nabla\tilde{\Phi}$ 
is assumed to be negligible compared to the mirror 
force $a_{\mathrm{m}}=-\mu\mathbf{b}\cdot\nabla B$.
The ratio between these two is of order,

\begin{equation}
  \label{accel_limit}
  \frac{a_{\|}}{a_{\mathrm{m}}}\sim\frac{Z e\tilde{\Phi}}{T}\frac{B}{\Delta B},
\end{equation}
with $\Delta B$ the typical amplitude of the helical magnetic wells.
Again, the proportionality to $Z$ makes it necessary to account for
$a_{\|}$ if the impurity transport is to be quantitatively \red{obtained}, 
although $\tilde{\Phi}$ can be sufficiently small for the bulk species.\\
On the other hand, the inclusion of $\tilde{\Phi}$ into the problem
makes the kinetic energy of the impurities
vary enough to violate the traditional neoclassical mono-energetic 
assumption that the velocity is nearly constant along the collisionless
particle orbits.\\
In the present work we put the focus on the computation of
the neoclassical particle flux of impurities including 
$\tilde{\Phi}$, thus abandoning the mono-energetic assumption, 
but keeping the radially local one. This latter approximation assumes
that the drifts across the flux surface are sufficiently small to treat them
perturbatively.
The calculations are performed with the Monte Carlo $\delta f$ PIC 
code \texttt{EUTERPE}. A concise description of it is 
given in section \ref{sec:euterpe}, highlighting the truncation 
of the global characteristics to perform local neoclassical runs.
Section \ref{sec:imps_with_phi1theta} shows particle flux calculations for 
C$^{6+}$, Ne$^{8+}$ and Fe$^{20+}$ in the LHD standard configuration 
including the poloidal variation of $\tilde{\Phi}(\theta)$,
with $\theta$ the poloidal coordinate. 
This is obtained from the solution of the steady-state ripple averaged
drift kinetic equation obtained by the $\texttt{GSRAKE}$ code \cite{Beidler_ppcf_43_2001}.
Section \ref{sec:imps_with_phi1sc} addresses
the calculation of $\tilde{\Phi}$ by \texttt{EUTERPE},
discussing a preliminary result that includes the dependence
of $\tilde{\Phi}$ on the toroidal coordinate $\phi$ and time $t$.
Finally, in section \ref{sec:discussion} a summary of the results 
and a discussion are presented.

\section{The \texttt{EUTERPE} code in the local neoclassical limit}
\label{sec:euterpe}

\texttt{EUTERPE} is a global $\delta f$ 
PIC Monte Carlo code, full-radius and full-flux surface, 
initially conceived for numerical simulation of linear gyro-kinetic 
micro-turbulence in 3D equilibria
\cite{Jost_pop_8_2001,Kornilov_nf_45.4_2005,Kleiber_aipproc_871_2006}.
After undergoing successive updates
the current version is non-linear, can treat multiple kinetic species simultaneously,
and perform electro-magnetic simulations. 
Recently it has been extended to include 
pitch angle scattering collisions \cite{Kauffmann_jpcs_260.1_2010}.\\
The set of phase space coordinates that \texttt{EUTERPE} uses is $\mathbf{z}=\{\mathbf{R},v_{\|},\mu\}$.
$\mathbf{R}$ is the position of the guiding center of the particle in neoclassical runs, 
or its gyro-center in gyro-kinetic ones.
In gyro-kinetic runs, the collisionless trajectory of a particle in phase space
in the electrostatic limit, for simplicity written in the vacuum 
approximation ($\nabla\times\mathbf{B}\rightarrow 0$), 
is determined by the following set of equations:

\begin{eqnarray}
&&\dot{\mathbf{R}}=\mathbf{v}_{\|}+\mathbf{v}_{E}+ \mathbf{v}_{\textrm{d}},\label{rdot}\\
&&\dot{v}_{\|}=\frac{q}{m}\mathbf{b}\cdot \mathbf{E}-\mu\mathbf{b}\cdot\nabla B
+\frac{v_{\|}}{B^2}\left(\mathbf{b}\times\nabla B\right)\cdot\mathbf{E},\label{vpardot}\\
&&\dot{\mu}=0\label{mudot},
\end{eqnarray}
where a dot above a symbol denotes a time derivative,
$\mathbf{E}=-\nabla\Phi$ is the electric field,
$\mathbf{v}_{E}=\mathbf{E}\times\mathbf{b}/B$ and 
$\Phi$ the electrostatic potential obtained from the gyro-kinetic quasi-neutrality
equation with arbitrary spatial dependence. 
The departure of the distribution function from the lowest order part $f_{0}$, 
$\delta f(\mathbf{z},t)$, is simulated using markers, whose 
trajectories are pushed according to eqs.~(\ref{rdot})-(\ref{mudot}),periodically interrupted by the application of a random change of their pitch angle
to account for collisions \cite{Takizuca_jcp_25.3_1977}.\\
It is straightforward to prove that eqs.~(\ref{rdot})-(\ref{mudot}) conserve the total 
energy of the particle and preserve the incompressibility of the 
phase space flow, i.e.  $\dot{\mathcal{E}}=0$ with $\mathcal{E}=v_{\|}^{2}/2+\mu B + (q/m)\Phi$
and $\nabla_{\mathbf{z}}\cdot \left(\mathcal{J}\dot{\mathbf{z}}\right)=0$ with $\nabla_{\mathbf{z}}=\left(\nabla,\partial_{v_{\|}},\partial_{\mu}\right)$ and $\mathcal{J}=B$ the Jacobian of our transformation.\\
The trajectory in real space followed by a particle 
given by eq.~(\ref{rdot}) is referred to as \textit{global} since
the radial magnetic and $\mathbf{E}\times \mathbf{B}$ 
drifts across the flux surface, 
$\mathbf{v}_{\mathrm{d}}\cdot\nabla s$ and
$\mathbf{v}_{E}\cdot\nabla s$ respectively, are accounted for. 
In contrast, the guiding center trajectory without these drifts
in lowest order is called \textit{local}. 
Local trajectories find routine application in neoclassical simulations 
and follow from assuming $\Phi\approx\Phi_{0}$, resulting in
$\mathbf{v}_{E}$ laying on the flux surface, and
$v_{\mathrm{d}}/v_{\|}\sim\delta$. This avoids the introduction 
of a radial derivative in $\partial \delta f /\partial s$ into 
the kinetic equation, which increases the dimensionality of the problem.
In this approximation the total energy is not conserved since the neglect
of $\mathbf{v}_{\mathrm{d}}$ unbalances a 
cancellation in $\dot{\mathcal{E}}$ resulting in 
$\dot{\mathcal{E}}=-(q/m)\mathbf{v}_{\mathrm{d}}\cdot\nabla\Phi_{0}$.
For impurities we need to retain $\tilde{\Phi}$ though,
and assume that $v_{\tilde{\Phi}}/v_{\|}\sim\delta$ to preserve the local \textit{ansatz}.
This and the approximation 
$\mathbf{b}\times\nabla B\cdot\nabla\Phi\approx \mathbf{b}\times\nabla B\cdot\nabla\Phi_{0}$  
brings additional non-conservation (of comparable
magnitude) associated with $\tilde{\Phi}$: 
$\dot{\mathcal{E}}=-(q/m)\mathbf{v}_{\mathrm{d}}\cdot\nabla\Phi_{0} - 
(q/m)\mathbf{v}_{\tilde{\Phi}}\cdot\nabla\Phi_{0}$ and 
$\dot{\mathcal{E}}_{0}=-(q/m)\mathbf{v}_{\mathrm{d}}\cdot\nabla\Phi_{0} - 
(q/m)v_{\|}\mathbf{b}\cdot\nabla\tilde{\Phi}$ where $\mathcal{E}_{0}=v_{\|}^{2}/2+\mu B + (q/m)\Phi_{0}$.
%
%
The variation of the energy during one collisional time $\tau$ can be 
estimated as $(\Delta\mathcal{E})_{\tau}/T\lesssim
(q\Phi_{0}/T)/(\Delta r/a)$, with 
$\Delta r\lesssim \delta v_{\|} \tau$ the scale of the total radial drift
in a typical collisional time, and $\nabla\sim a^{-1}$.\\
Our equations for the marker trajectory in phase space for a local neoclassical 
run are thus,

\begin{figure}
  \begin{center}
    \includegraphics[width=0.31\textwidth,angle=0]
    {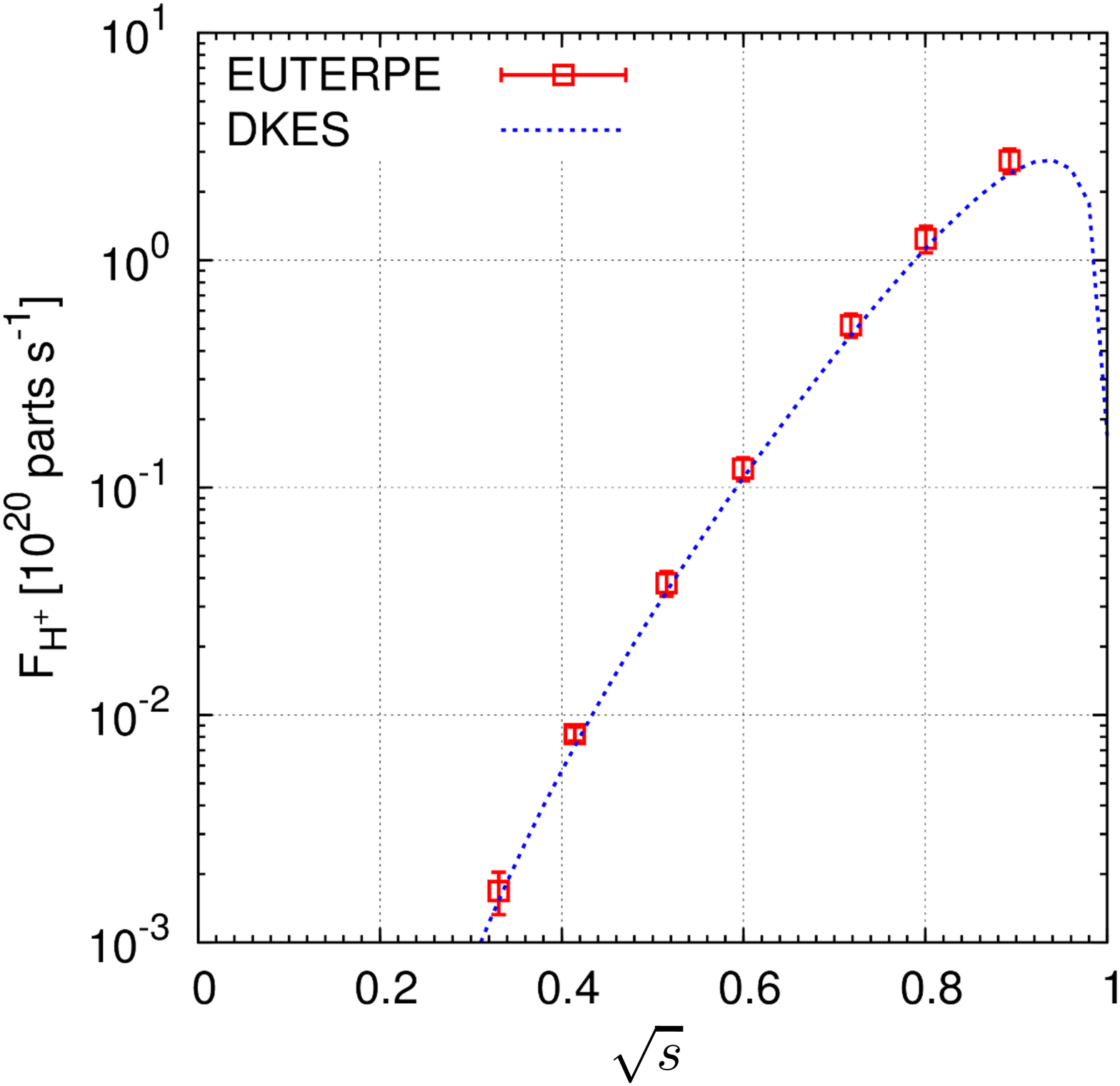}
    \includegraphics[width=0.31\textwidth,angle=0]
    {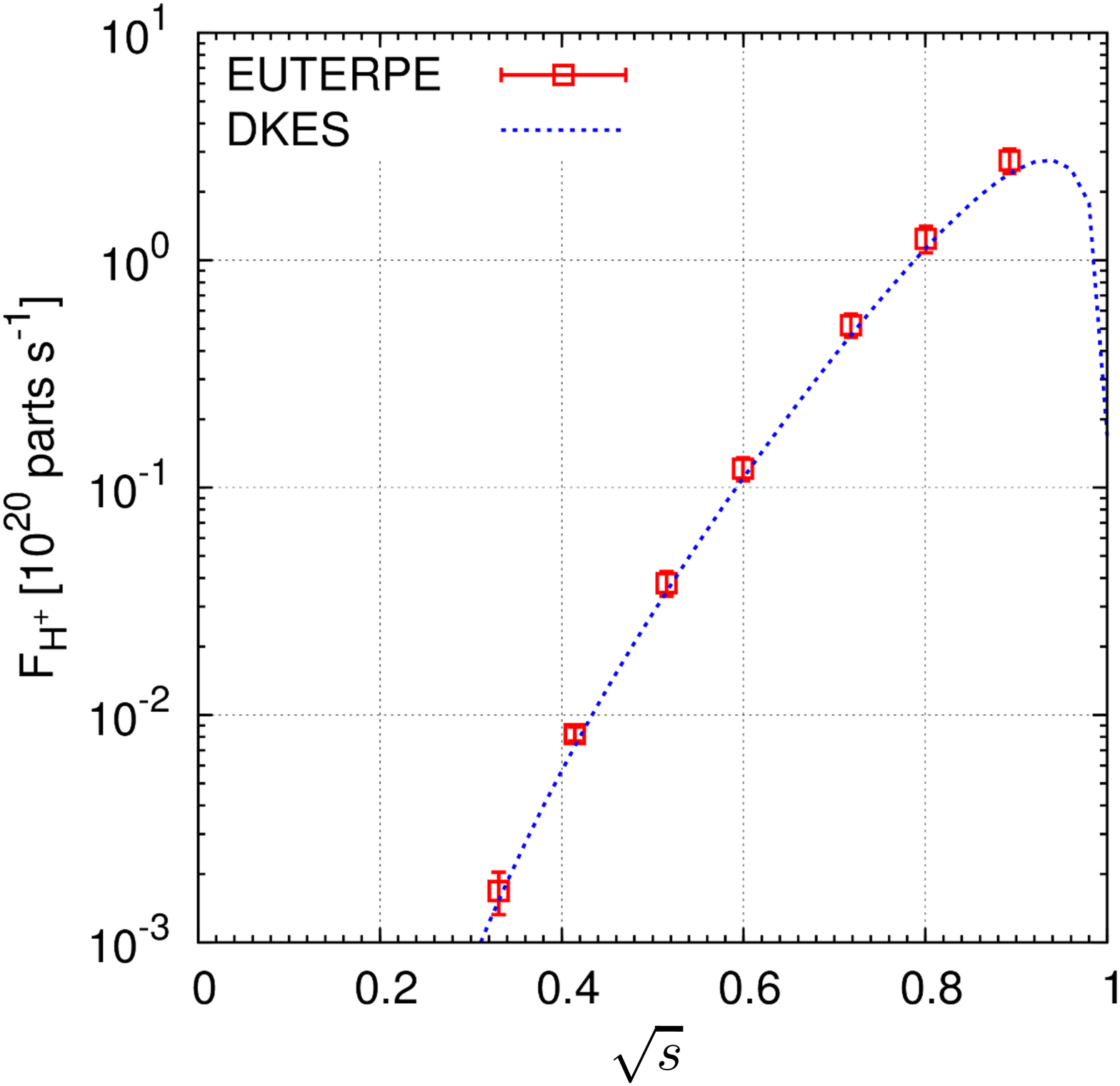}
    \includegraphics[width=0.33\textwidth,angle=0]
    {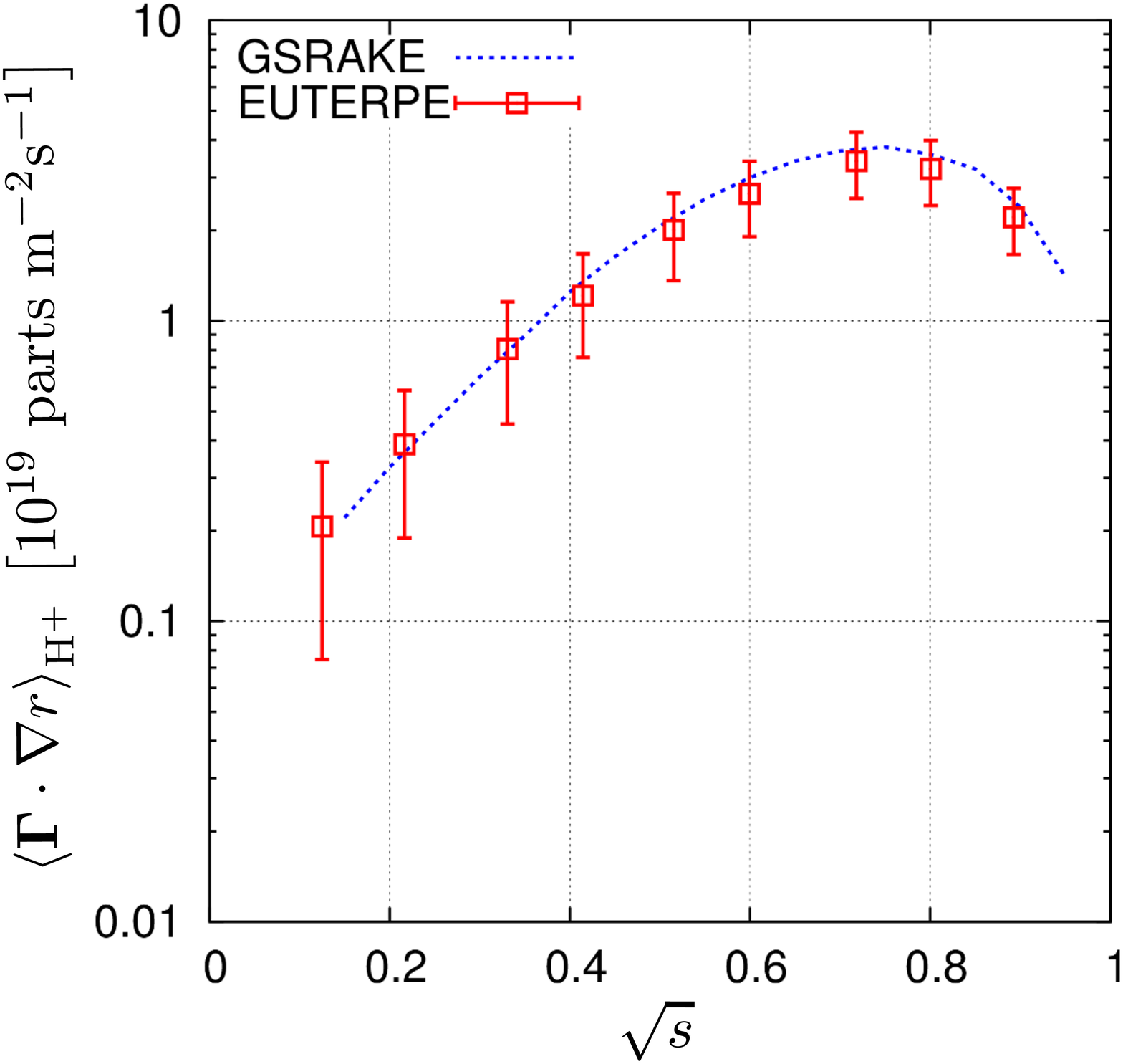}
    \caption{Particle flux of hydrogen in 
an axisymmetric circular cross section tokamak for the
radial density flat profile of $n=10^{20}$ m$^{-3}$ and 
a temperature profile $T=2(1-s)$ keV (left), 
and for $n=(1-s^4)\times 10^{19}$ m$^{-3}$ at 
a fixed temperature of $T=2$ keV (center). 
In the right, particle flux density 
of hydrogen in LHD for the profiles \red{$n=(6-5s^4)\times 10^{19}$ m$^{-3}$ and $T=(2-1.8s)$} keV including $E_{r}$.}
    \label{fig:benchmark}
  \end{center}
\end{figure}

\begin{eqnarray}
&&\dot{\mathbf{R}}=\mathbf{v}_{\|}+\mathbf{v}_{E0},\label{rdot_local}\\
&&\dot{v}_{\|}=-\frac{q}{m}\mathbf{b}\cdot\nabla\tilde{\Phi}-\mu\mathbf{b}\cdot\nabla B+
\frac{v_{\|}}{B^2}\left(\mathbf{b}\times\nabla B\right)\cdot\mathbf{E}_{r},\label{vpardot_local}\\
&&\dot{\mu}=0\label{mudot_local},
\end{eqnarray}
with $\mathbf{v}_{E0}=\mathbf{E}_{r}\times\mathbf{b}/B$. 
For the tokamak, where 
the small-$\mathbf{E}_{r}$ limit $v_{E0}/v_{\|}\sim \delta$ applies \cite{Helander_book} 
(unless the plasma rotates rapidly)
the second term in eq.~(\ref{rdot_local}) and the third in eq.~(\ref{vpardot_local}) are
neglected. Assuming that $f_{0}$ is a Maxwellian distribution function 
$f_{0}=f_{\mathrm{M}}(s,v_{\|},v_{\bot})=\left[n_{0}(s)/(2\pi)^{3/2} v_{\mathrm{th}}^{3}(s)\right]\exp\left[-\left(v_{\|}^{2}+v_{\bot}^2\right)/2v_{\mathrm{th}}^{2}(s)\right]$, with 
$v_{\mathrm{th}}=\sqrt{T/m}$ and $T=T(s)$ the temperature, 
the resulting kinetic equation  reads as follows:

\begin{equation}
\label{local_dke}
\begin{split}
\frac{\partial\delta f}{\partial t} +
\dot{\mathbf{R}}\cdot\nabla\delta f +
\dot{v}_{\|}\frac{\partial \delta f}{\partial v_{\|}}=&
-f_{\mathrm{M}}\left[\frac{1}{n}\frac{\partial n}{\partial s}+
\left(\frac{mv^{2}}{2T}-\frac{3}{2}\right)
\frac{1}{T}\frac{\partial T}{\partial s}\right]
\left(\mathbf{v}_{\mathrm{d}}+\mathbf{v}_{\tilde{\Phi}}\right)\cdot\nabla s+\\
&+\frac{q}{m}\frac{f_{\mathrm{M}}}{v_{\mathrm{th}}^{2}}
\left(\mathbf{v}_{\|}+\mathbf{v}_{\mathrm{d}}\right)\cdot
\left(\mathbf{E}_{r}-\nabla\tilde{\Phi}\right).
\end{split}
\end{equation}
Note that in obtaining eq. (\ref{local_dke}) the original drift, eq. (\ref{rdot}), 
was considered and those
terms higher than first order in $\delta$  were truncated.
Moreover since the approximation $\tilde{\Phi}\ll \Phi_{0}$
still holds the radial electric field related to $\tilde{\Phi}$ is
considered negligible compared to $\mathbf{E}_{r}$.
\texttt{EUTERPE}, in the present neoclassical modality, 
was satisfactorily benchmarked against theory and the codes
\texttt{DKES} and \texttt{GSRAKE}.
Figure \ref{fig:benchmark} shows a comparison between 
the particle fluxes obtained with these codes in different cases.
It is important to remark that in this comparison $\tilde{\Phi}$ 
is missing since \texttt{DKES} and \texttt{GSRAKE} assume mono-energetic
trajectories.

\section{Impurity particle transport in the presence of $\tilde{\Phi}\left(\theta\right)$}
\label{sec:imps_with_phi1theta}

The code \texttt{GSRAKE} solves the ripple-averaged drift 
kinetic equation providing, apart from the neoclassical fluxes,
the first order corrections to the equilibrium density and electrostatic potential 
$\tilde{n}\left(r,\theta\right)$ and $\tilde{\Phi}\left(r,\theta\right)$.
The magnetic configuration is accounted for by means of 
a multiple-helicity model and the bounce average is performed
along the toroidal coordinate $\phi$. Moments of the 
perturbed distribution function and the flux-surface-average are defined in the 
\textit{reduced} phase-space, in which the toroidal angle coordinate has been
eliminated by performing the ripple average. This process \textit{removes} 
the $\phi$ dependence of $B$ appearing in the Jacobian of the initial phase-space.
The resulting $\theta$-dependent 
$\tilde{n}$ and $\tilde{\Phi}$ are
written in Fourier series, and their coefficients are iteratively adjusted 
to simultaneously fulfill the 
quasi-neutrality and ambipolarity constraints among bulk ions and
electrons (for the details 
of the code see refs.~\cite{Beidler_ppcf_43_2001, Beidler_isw_2005}).\\

\begin{figure}[t]
  \begin{center}
    \begin{subfigure}[b]{0.3\textwidth}
      \centering
      \includegraphics[width=\textwidth,angle=0]
      {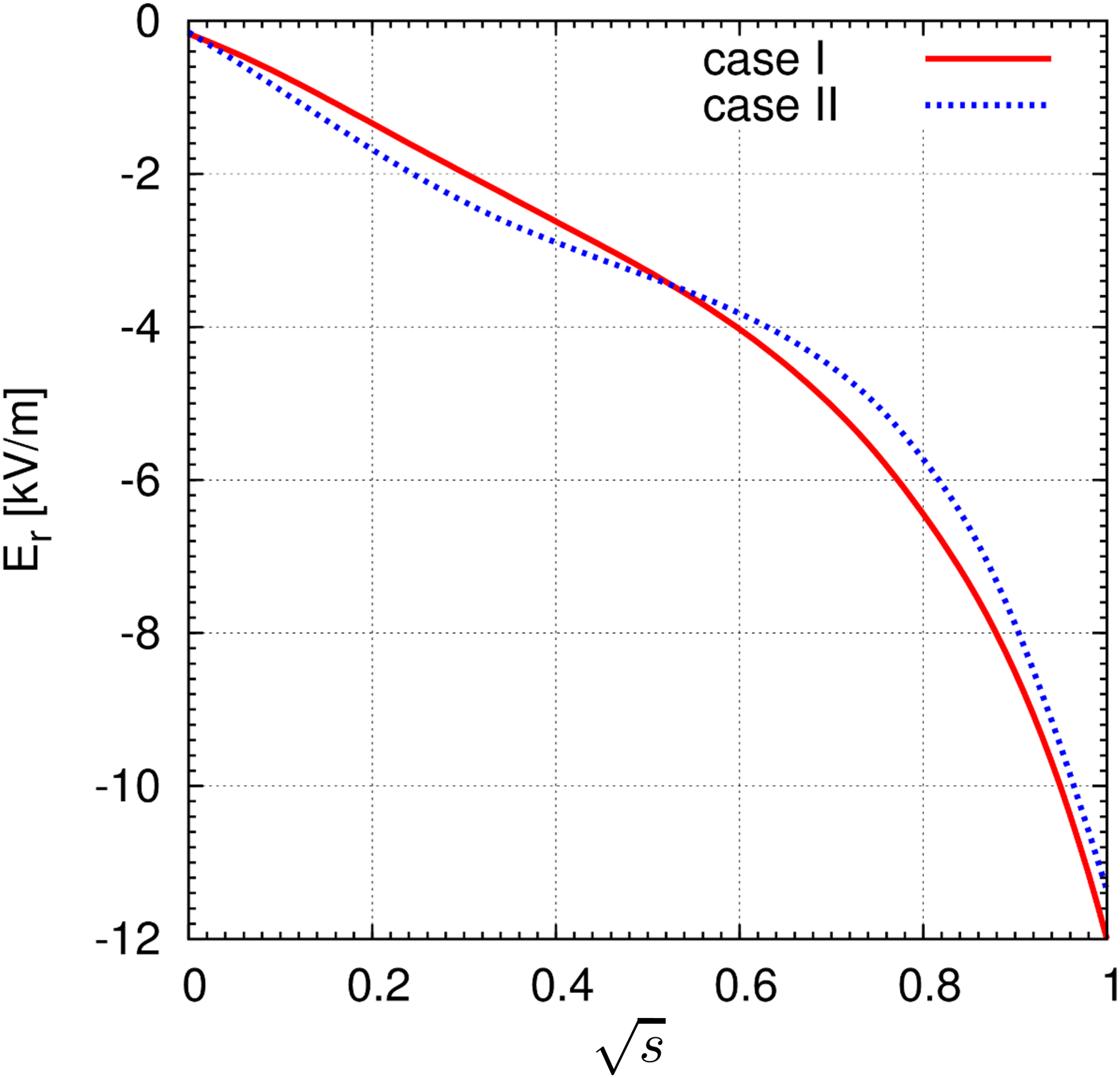}
      \caption{(a)}
    \end{subfigure}
    \begin{subfigure}[b]{0.33\textwidth}
      \centering
      \includegraphics[width=\textwidth,angle=0]
      {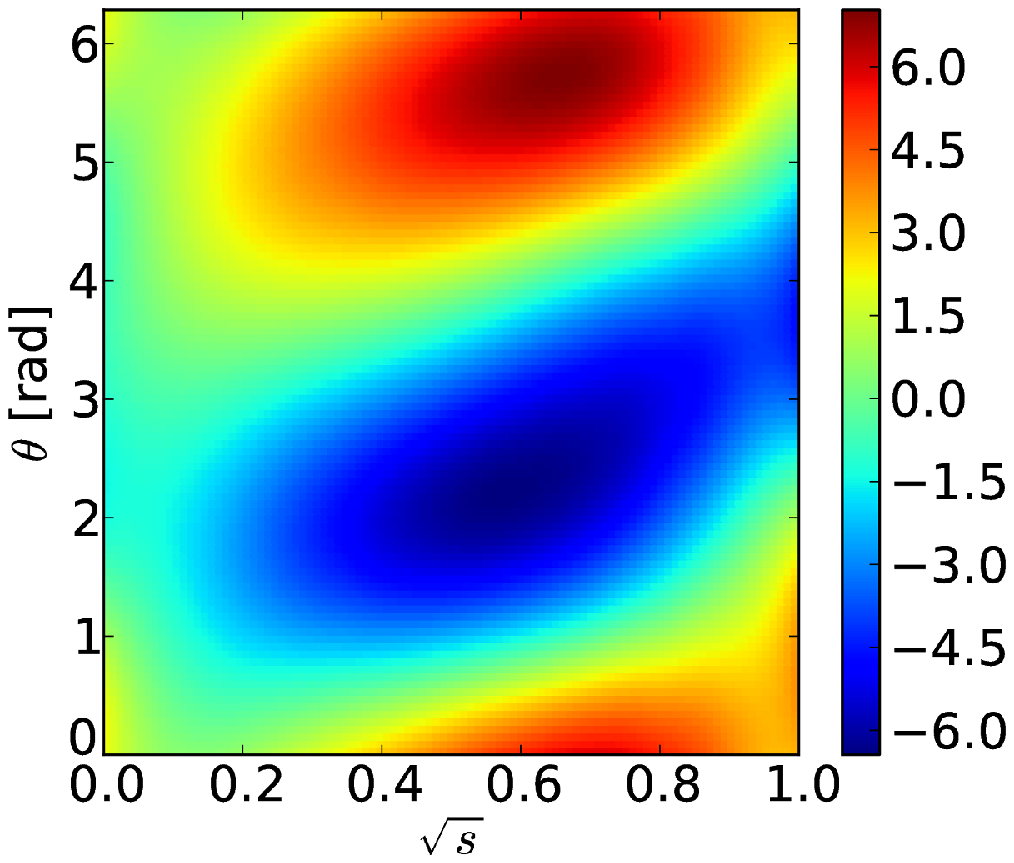}
      \caption{(b)}
    \end{subfigure}
    \begin{subfigure}[b]{0.33\textwidth}
      \includegraphics[width=\textwidth,angle=0]
      {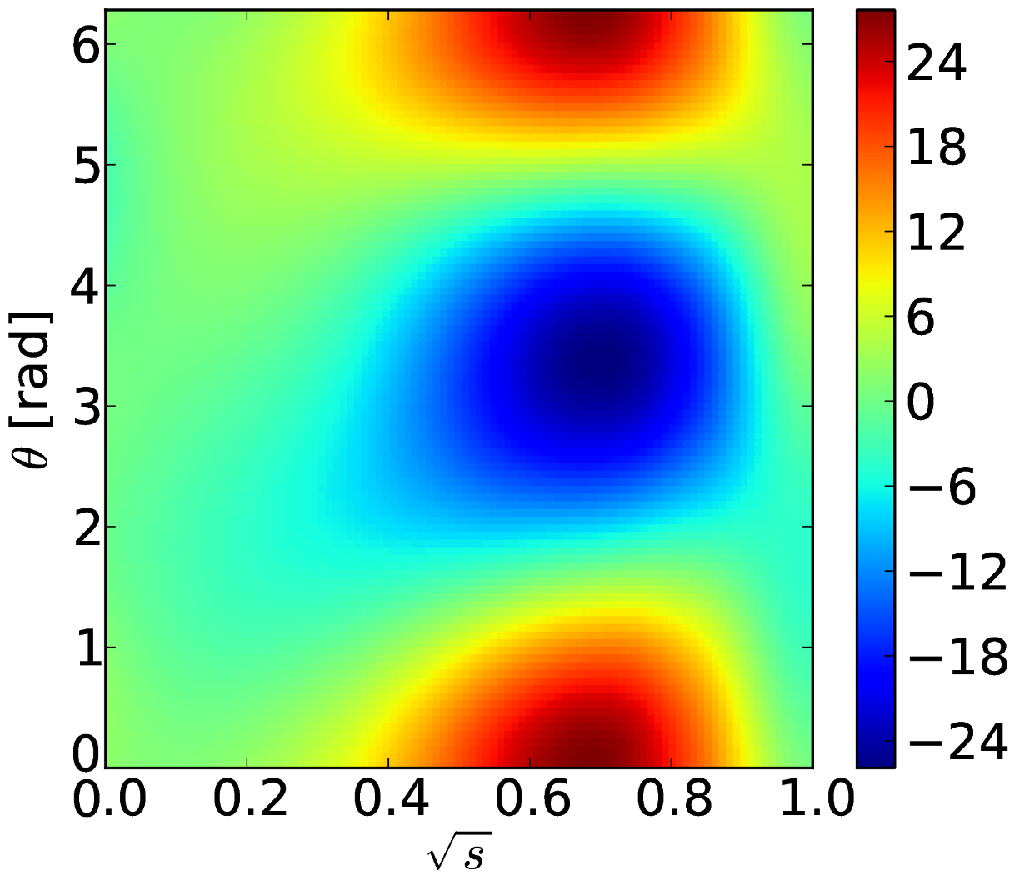}
      \caption{(c)}
    \end{subfigure}
    \caption{(a) Ambipolar radial electric field obtained with \texttt{GSRAKE}
      in the two cases considered for the impurity runs launched by \texttt{EUTERPE}.
      The corresponding $\tilde{\Phi}$ are represented in plot (b) for case I,
      and (c) for case II.}\label{fig:gsrake_cases}
  \end{center}
\end{figure}

In the present section
the $E_{r}$ and $\tilde{\Phi}$ obtained with \texttt{GSRAKE} for
two different sets of density and temperature profiles 
have been used as input for \texttt{EUTERPE}, which in turn
obtains the particle flux density $\left<\boldsymbol{\Gamma}\cdot\nabla r\right>$. 
This particular choice 
of tasks is due to the fact that \texttt{EUTERPE} is capable to integrate
the impurity trajectories in the 4D phase space resulting from the breakdown
of the mono-energetic approximation for impurities while $\texttt{GSRAKE}$ 
is not. On the other hand, due to the radial locality of the trajectories 
$E_{r}$ cannot be computed by \texttt{EUTERPE} unless that this is performed
iteratively, which adds an unnecessary computational cost due the slow convergence
of the fluxes for the bulk species.
The magnetic equilibrium considered corresponds to the
standard LHD configuration with $R\approx 3.73$ m, $a\approx 0.6$ m and
$B_{00}(\sqrt{s}=0.5)=2.53$ T.
Two cases have been considered and are
labeled as \textit{case I} and \textit{case II}.
The equilibrium temperature and density 
profiles considered in the former case for the bulk ions (hydrogen) are
$n=(6-5s^4)\times 10^{19}$ m$^{-3}$ and 
$T=(2-1.8 s)$ keV.
The rather lower collisional case II
considers instead
$n=5(1-0.2s^4)\times 10^{19}$ and 
$T=3(1-0.93s)$.
The temperature profiles are taken to be the 
same for electrons and impurities, and their 
central density \red{are adjusted} to fulfill quasi-neutrality.
The effective charge is set to $Z_{\mathrm{eff}}=1.05$ at all radii.
In fig. \ref{fig:gsrake_cases}(a) the ambipolar radial electric field 
is displayed for both pairs of profiles, in fig.~\ref{fig:gsrake_cases}(b)
the $\tilde{\Phi}$ map dependent on $\sqrt{s}$ and $\theta$
for case I is given, while \ref{fig:gsrake_cases}(c) represents the one 
corresponding to the case II.\\

\begin{figure}[t]
  \begin{center}
    \begin{subfigure}[b]{0.3\textwidth}
      \centering
      \includegraphics[width=\textwidth,angle=0]
      {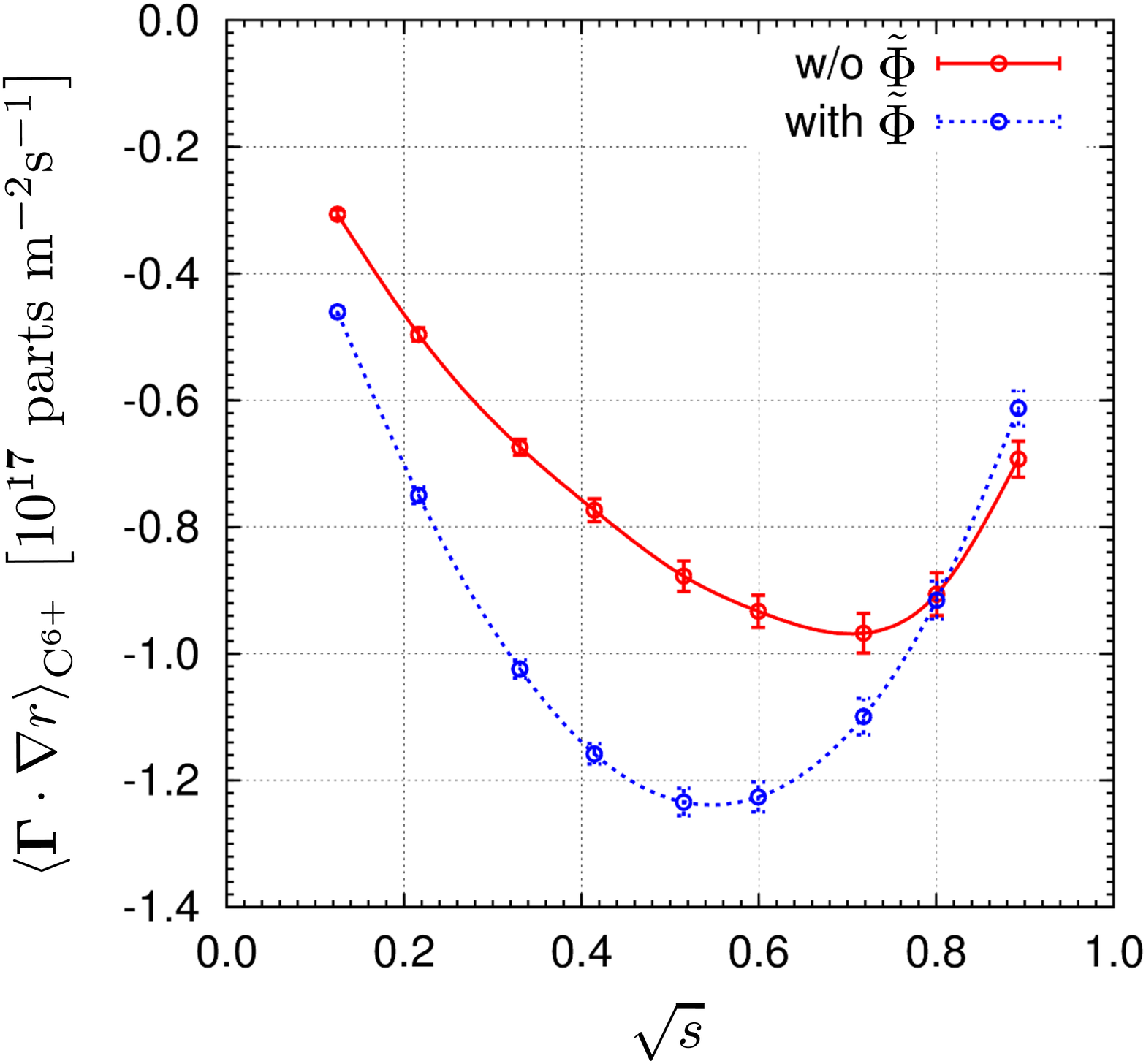}
      \caption{(a)}
    \end{subfigure}
    \begin{subfigure}[b]{0.3\textwidth}
      \centering
      \includegraphics[width=\textwidth,angle=0]
      {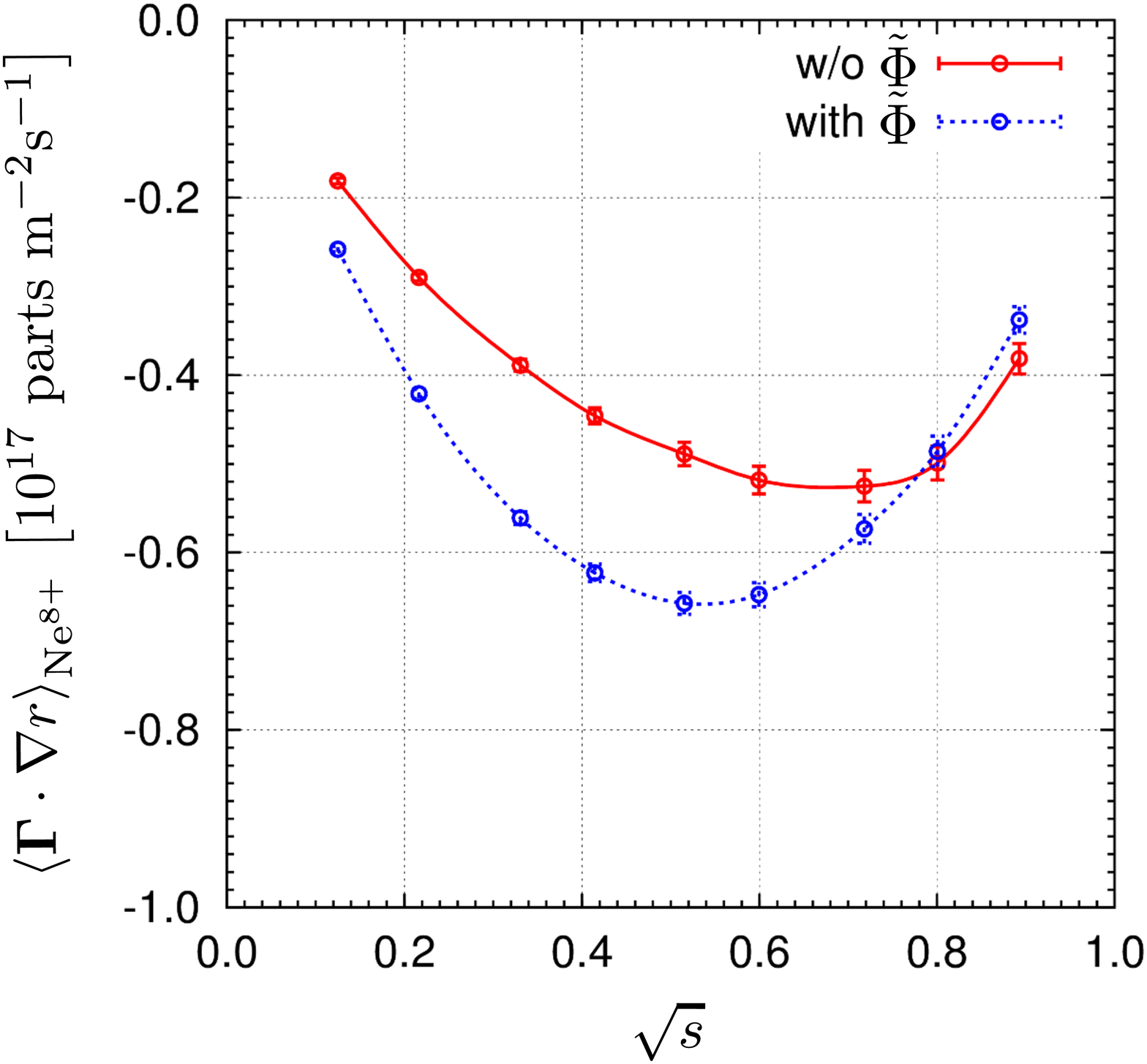}
      \caption{(b)}
    \end{subfigure}
    \begin{subfigure}[b]{0.3\textwidth}
      \centering
      \includegraphics[width=\textwidth,angle=0]
      {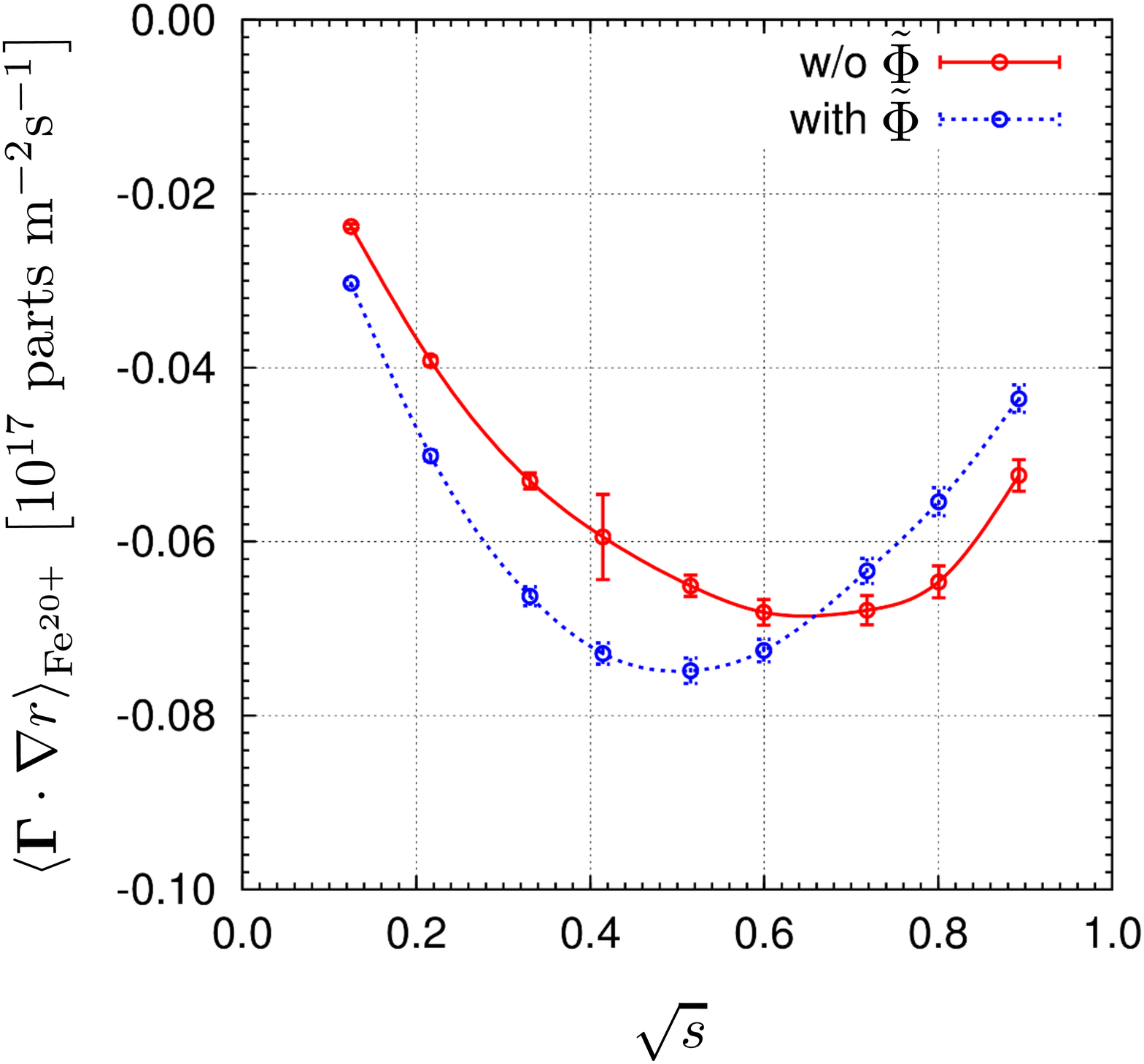}
      \caption{(c)}
    \end{subfigure}
    \caption{Particle flux density of C$^{6+}$ (a), Ne$^{8+}$ (b) and  Fe$^{20+}$ (c) as a function of $\sqrt{s}$ in LHD for the set of profiles considered in case I (see text).}\label{fig:imps_with_gsrake1}
  \end{center}
\end{figure}

\begin{figure}[t]
  \begin{center}
    \begin{subfigure}[b]{0.3\textwidth}
      \centering
      \includegraphics[width=\textwidth,angle=0]
      {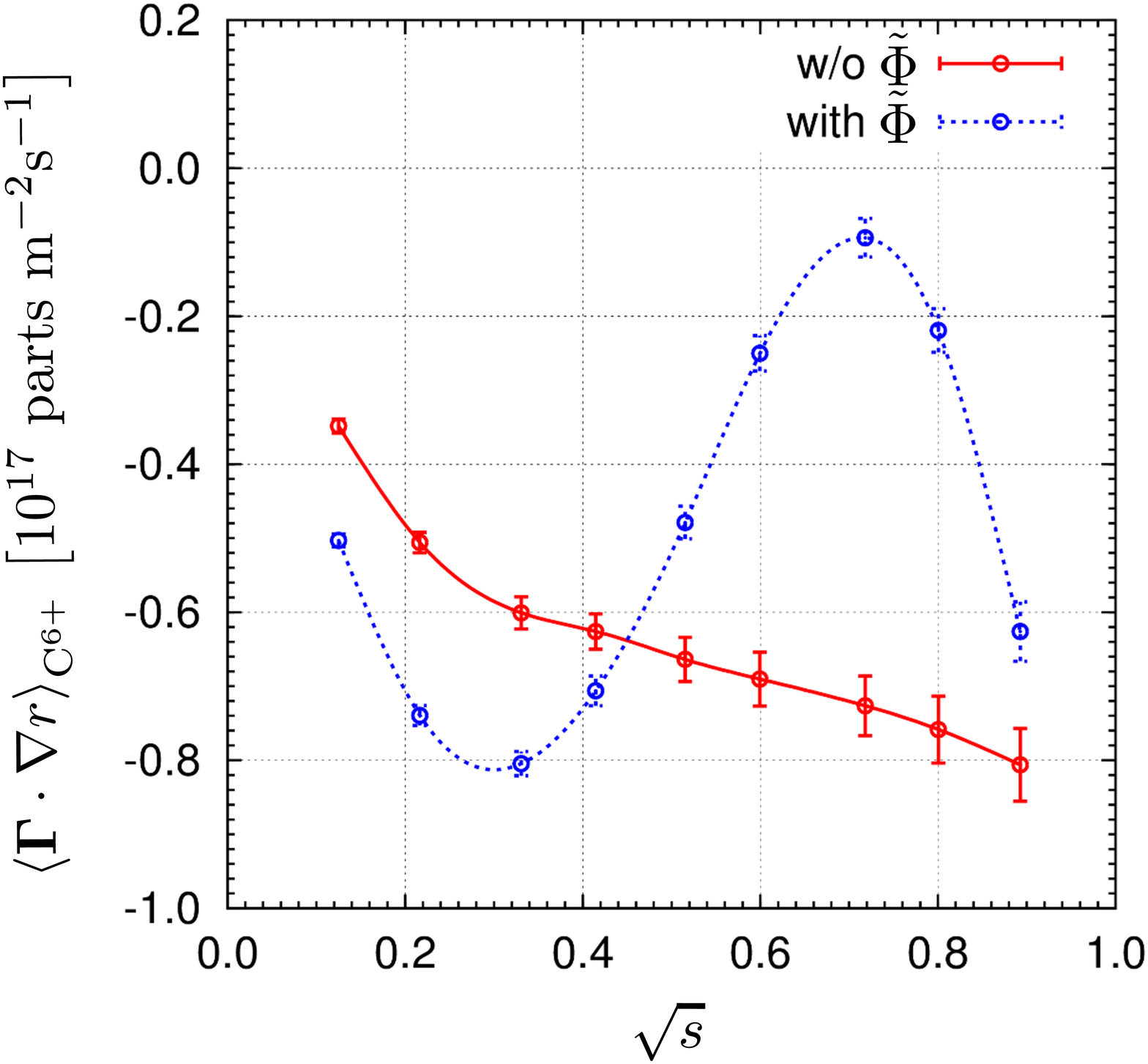}
      \caption{\red{(a)}}
    \end{subfigure}
    \begin{subfigure}[b]{0.3\textwidth}
      \centering
      \includegraphics[width=\textwidth,angle=0]
      {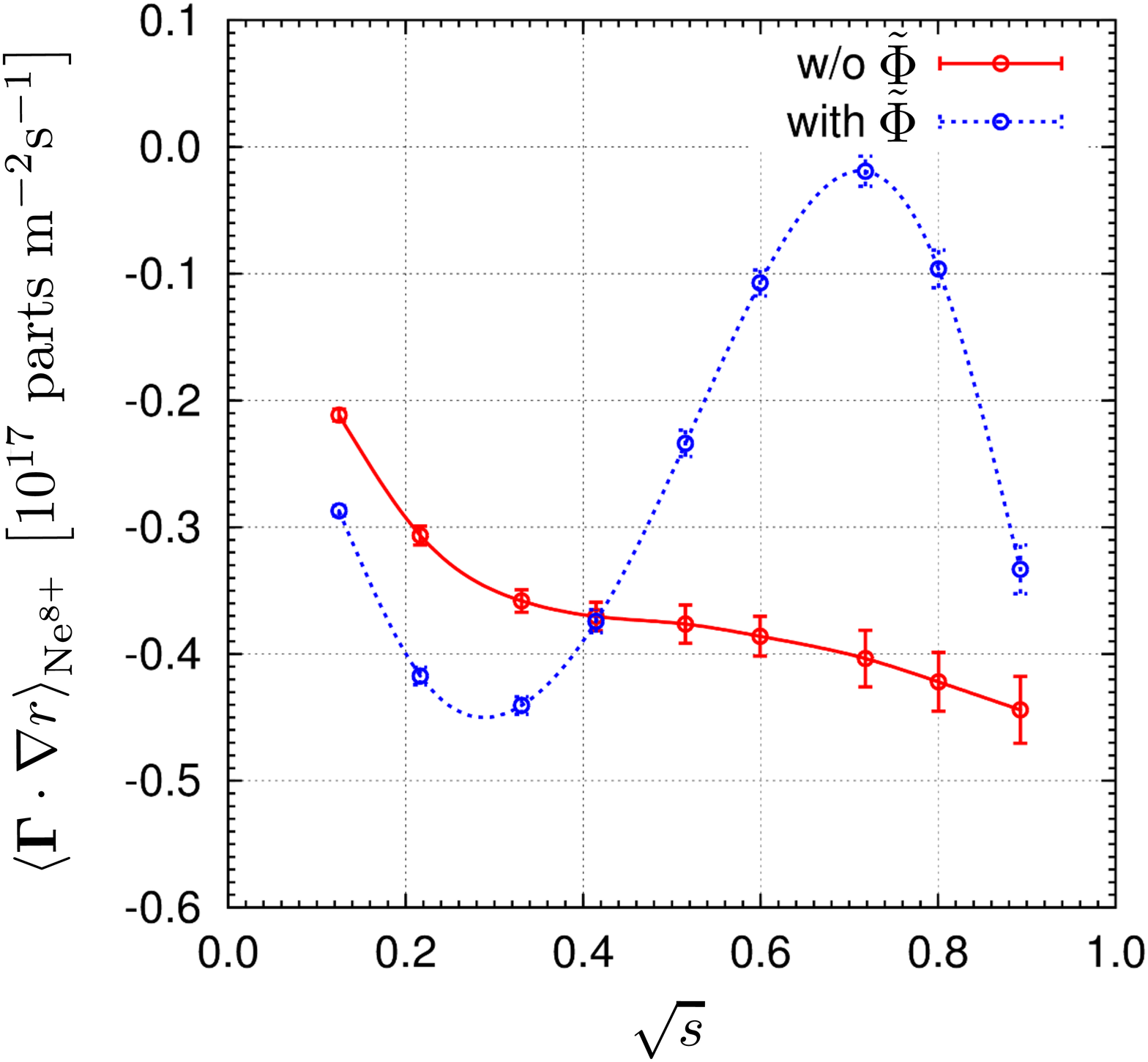}
      \caption{\red{(b)}}
    \end{subfigure}
    \begin{subfigure}[b]{0.3\textwidth}
      \centering
      \includegraphics[width=\textwidth,angle=0]
      {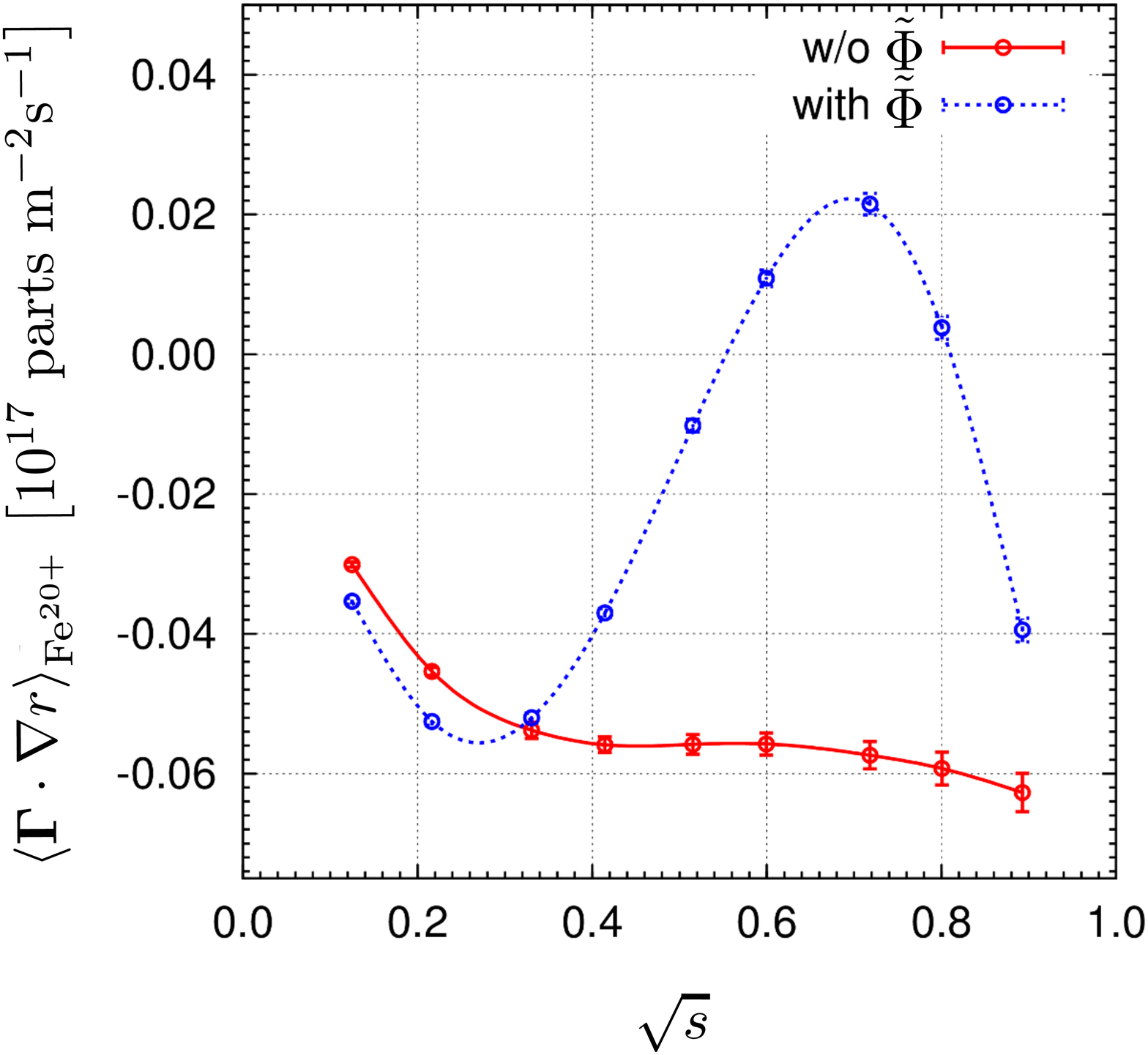}
      \caption{\red{(c)}}
    \end{subfigure}
    \caption{Particle flux density of C$^{6+}$ (a), Ne$^{8+}$ (b) and  Fe$^{20+}$ (c) as a function of $\sqrt{s}$ in LHD for the set of profiles considered in case II (see text).}\label{fig:imps_with_gsrake2}
  \end{center}
\end{figure}

\begin{figure}[t]
  \begin{center}
      \centering
      \includegraphics[width=0.4\textwidth,angle=0]
      {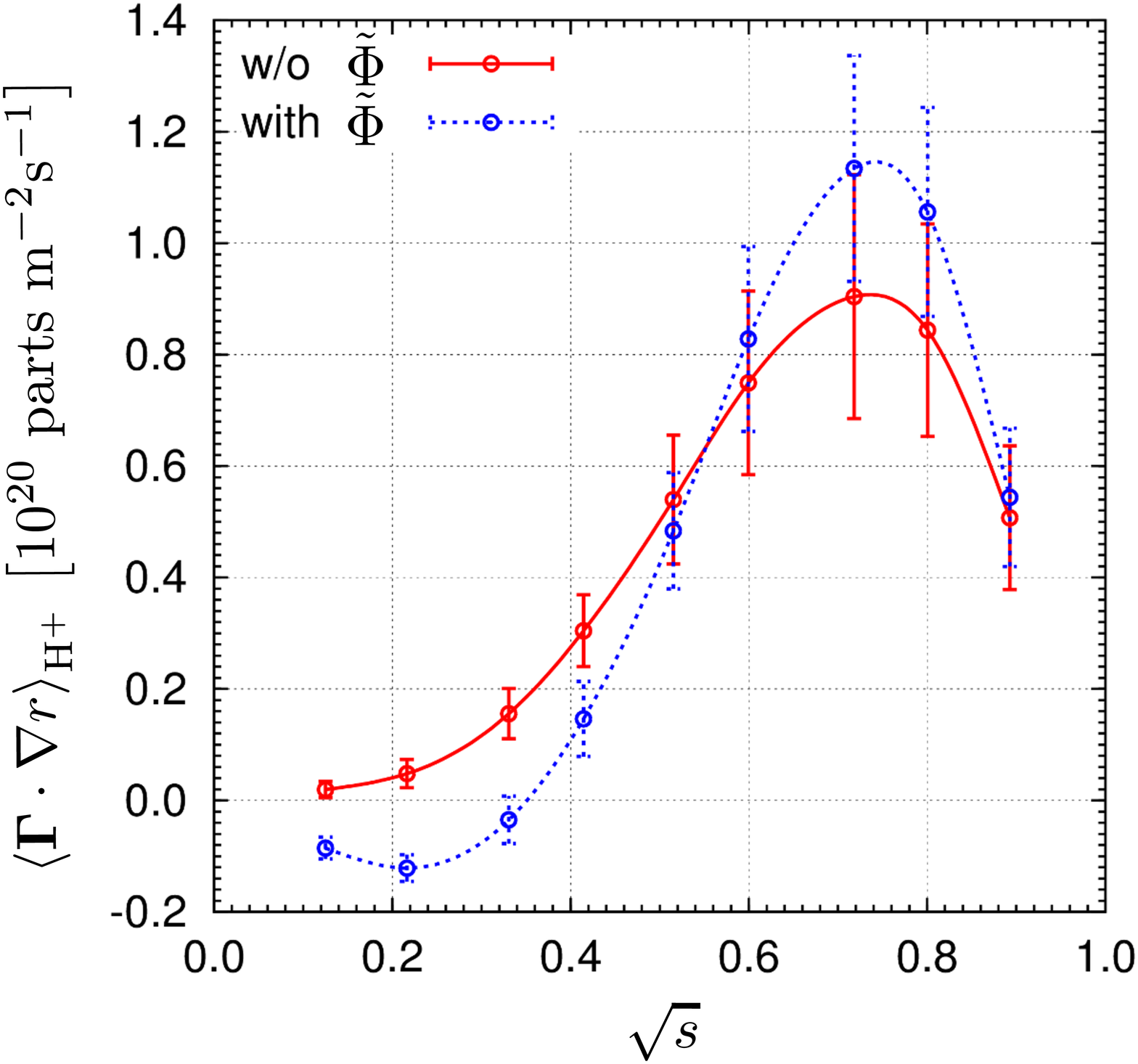}
    \caption{Particle flux density of H$^{+}$ as a function of $\sqrt{s}$ in LHD for the case II profiles (see text).}\label{fig:h1_with_gsrake3}
  \end{center}
\end{figure}

The simulations run with \texttt{EUTERPE} were carried out for
C$^{6+}$, Ne$^{8+}$ and Fe$^{20+}$.
The radial profiles of the particle flux density are shown 
in figs.~\ref{fig:imps_with_gsrake1}(a)-(c) for case I.
In this set of plots
the impact of $\tilde{\Phi}$ on the three species is detrimental
up to $\sqrt{s}\approx 0.7-0.8$
from the impurity accumulation perspective. 
From that radial location outwards the trend 
is reversed and the inward flux in the presence of $\tilde{\Phi}$
becomes weaker. We conclude that $\tilde{\Phi}$ can
act either to amplify or mitigate the inward flux driven by $\mathbf{E}_{r}$.
It can also be observed that $\tilde{\Phi}$ affects the C$^{6+}$ particle flux 
more than Fe$^{20+}$'s. At a first glance this may look like contradictory 
with what eqs. (\ref{drift_limit})
and (\ref{accel_limit}) suggest. But it is important to notice that what 
those expressions imply is that the 
acceleration due to $\tilde{\Phi}$ must be included in $\dot{v}_{\|}$ and 
that $\mathbf{v}_{\tilde{\Phi}}$ must be put on the same order than
$\mathbf{v}_{\text{\tiny{d}}}$. This in turn brings to the  
source of the kinetic equation, eq. (\ref{local_dke}), three more terms 
traditionally neglected containing $\tilde{\Phi}$. 
The balance of these terms with the rest and with each other
does not necessarily lead to a more noticeable impact on the transport 
produced by $\tilde{\Phi}$ with the increasing $Z$.
\\
The same behavior is shown for case II in figs.~\ref{fig:imps_with_gsrake2}(a)-(c).
Again $\tilde{\Phi}$ breaks the monotonic growth  
of the inward impurity flux with increasing $s$. 
The change in the trend is observed 
at a different radial location than in case I, 
$\sqrt{s}\approx 0.3-0.4$, which corresponds to 
almost the position where the electrostatic potential starts to be 
appreciably large, see figure \ref{fig:gsrake_cases}(c).
A remarkable feature in this case is that 
the particle flux approaches the value of zero at $\sqrt{s}\approx 0.7$ 
for C$^{6+}$ and Ne$^{8+}$ and is even positive for Fe$^{20+}$.
That radial position corresponds to the 
maximum amplitude of $\tilde{\Phi}$ along $\theta$.\\
At this point it is convenient to recall the discussion about the weight of $\tilde{\Phi}$
for driving radial transport and trapping particles
reflected by the expressions (\ref{drift_limit})
and (\ref{accel_limit}).
First, it is important to notice that in case I
the variation of $\tilde{\Phi}$ leads to an electrostatic energy 
variation of approximately $1$\% of the thermal energy for a unit charge.
In case II
the same variation would result in \red{approximately a $2$\% change}.
This indicates that the strong change in the behavior of the 
particle flux cannot be explained by such increase in the ratio $q\tilde{\Phi}/T$ exclusively.
This points out also to the non-trivial interplay between
all terms in the kinetic equation with and without $\tilde{\Phi}$.
This may lead to the natural question of how 
much the transport of hydrogen nuclei can be affected by $\tilde{\Phi}$.
In figure (\ref{fig:h1_with_gsrake3}) the comparison between the particle
flux density of H$^+$ with and without $\tilde{\Phi}$ for the case II is represented.
The profiles does not change qualitatively as in the case of impurities. Nevertheless two
important features should be emphasized. On the one hand the flux turns to be negative below $\sqrt{s}<0.3$ 
under the action of $\tilde{\Phi}$, and on the other the value 
at its maximum is enhanced in an approximately a 20 $\%$. These changes, although not as 
significant as in the cases with impurities, represent a noticeable effect.

\section{The self-consistent calculation of  $\tilde{\Phi}\left(\theta,\phi,t\right)$ in \texttt{EUTERPE}}
\label{sec:imps_with_phi1sc}

\begin{figure}[t]
  \begin{center}
    \begin{subfigure}[b]{0.4\textwidth}
      \centering
      \includegraphics[width=\textwidth,angle=0]
      {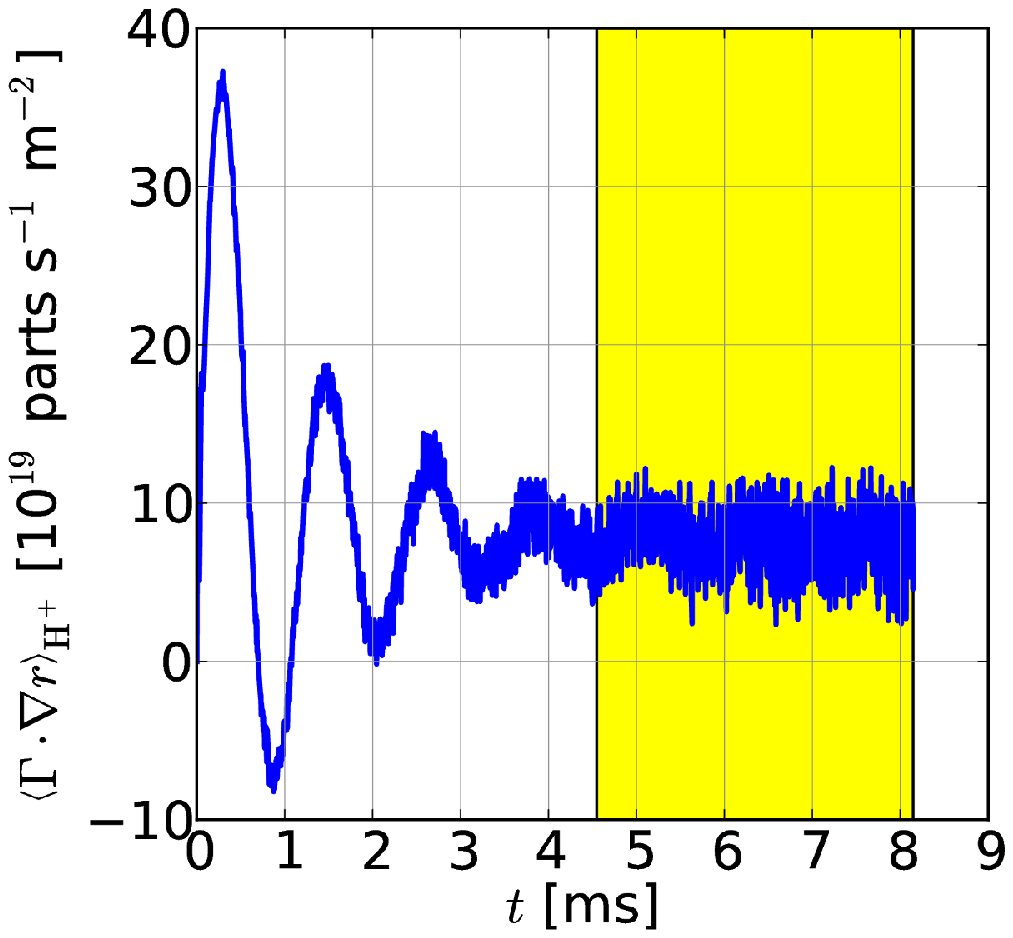}
      \caption{(a)}
    \end{subfigure}
    \begin{subfigure}[b]{0.415\textwidth}
      \centering
      \includegraphics[width=\textwidth,angle=0]
      {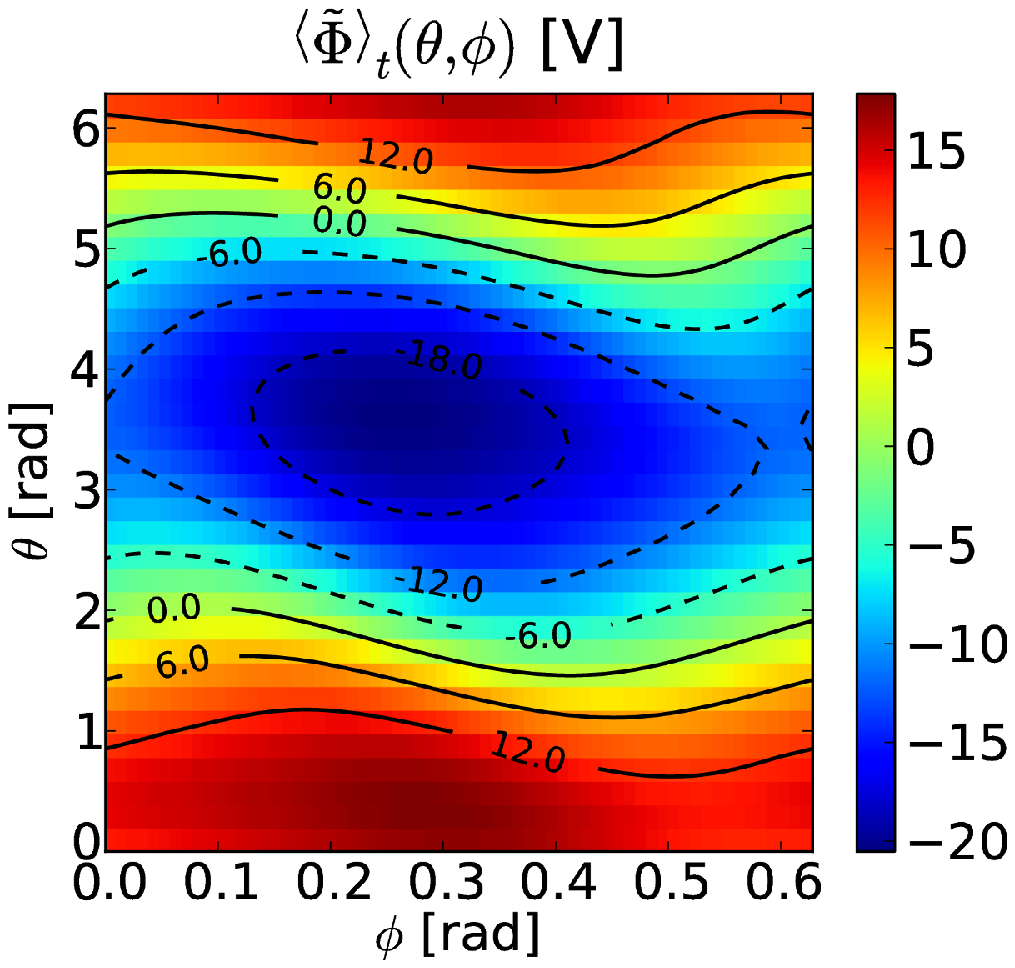}
      \caption{(b)}
    \end{subfigure}
    \caption{(a) Time dependent flux of hydrogen in LHD for the case II profiles 
      (see text in section \ref{sec:imps_with_phi1theta}) at $\sqrt{s}=0.5$. 
      (b) Time averaged electrostatic potential 
      $\langle\tilde{\Phi}\rangle_{t}(\theta,\phi)$.
    }
    \label{fig:phi2d_maps}
  \end{center}
\end{figure}

In light of these results it is of immediate interest to extend
the calculations to account for the toroidal dependence of $\tilde{\Phi}$.
Apart from the role that this can 
play, it is important to remember that 
the ripple average also reduces
the number of possible configurations that can be studied.
Although this is beyond the scope of the present work, we 
show in this section a preliminary result where
$\tilde{\Phi}(\theta,\phi,t)$ is obtained self-consistently with \texttt{EUTERPE}.
The approach rests on 
the fulfilment of the neutrality condition
on the density perturbations for each species, $\sum_{s}\tilde{n}_{s}=0$. 
The starting point is the 
quasi-neutrality gyro-kinetic equation \cite{Hahm_pf_31.9_2670_1988}
that the code solves at each time step.
Considering for simplicity kinetic ions and adiabatic electrons, 
the equation reads as follows:

\begin{equation}
\label{eq:poisson}
e\left<\tilde{n}_{i}\right>+m_i
\nabla\cdot\left(\frac{n_{0i}}{B^{2}}\nabla_{\bot}\tilde{\Phi}\right)-
\frac{e^{2}n_{0e}}{T_{e}}\left(\tilde{\Phi}-\bar{\Phi}\right)=0,
\end{equation}
where the lower indices $i$ and $e$ denote
bulk ions and electrons respectively, 
$n_{0}$ is the equilibrium density, 
\red{$\left<\tilde{n}_{i}\right>$} is the gyro-averaged ion density, 
$\bar{\Phi}$ is the flux-surface averaged electrostatic potential, and
$m_{i}$ the ion mass.
The equation is simplified by
invoking the limit $k_{\bot}\rho\ll 1$ 
with $k_{\bot}$ the perpendicular characteristic
wave-length of the fluctuations. Since the second
term in eq.~(\ref{eq:poisson}) is a factor 
$k_{\bot}^{2}\rho^{2}$ smaller than the others it
can be neglected. In addition, the gyro-average operation 
is unnecessary in drift kinetics. And finally, 
by flux surface averaging of the remaining expression,
it can be shown that $\bar{\Phi}=\tilde{\Phi}_{0,0}=0$.
The sub-indices in $\tilde{\Phi}$ denote that the poloidal and toroidal
mode numbers, $m$ and $n$ respectively, are both equal to zero.
The final expression is then:

\begin{equation}
\tilde{n}_{i}=\frac{e}{T_{e}}n_{0e}\tilde{\Phi}.
\end{equation}
In order to prevent the code from developing short wave-length 
unstable modes, only low mode numbers are retained by employing a
spectral filter set to $0\le m\le 4$ and $-4\le n \le 4$.
Since the trajectories are radially local, 
it is difficult to calculate $E_{r}$. An iterative adjustment
of $E_r$ until ambipolarity is fulfilled is not feasible 
due to the dynamic nature of the simulation, which makes it impossible 
to determine the flux before it has saturated. 
Thus, $E_r$ is imposed externally.\\
Figures \ref{fig:phi2d_maps}(a)-(b) show the result of a simulation
for $\tilde{\Phi}(\theta,\phi,t)$ at $\sqrt{s}=0.5$, 
considering the profiles of the labeled as case II in 
section \ref{sec:imps_with_phi1theta}.
The calculation of both the hydrogen flux density $\left<\boldsymbol{\Gamma}\cdot\nabla r\right>_{\mathrm{H}^+}$ 
and $\tilde{\Phi}$ 
is dynamical, as has already been mentioned. In figure \ref{fig:phi2d_maps}(a)
$\left<\boldsymbol{\Gamma}\cdot\nabla r\right>_{\mathrm{H}^+}$ is shown as a function of time. Once it has reached the
stationary value, a time average of the potential is performed.
The resulting time averaged potential $\langle\tilde{\Phi}\rangle_{t}(\theta,\phi)$
is represented in fig.~\ref{fig:phi2d_maps}(b). The time interval 
considered for the average is colored yellow in fig.~\ref{fig:phi2d_maps}(a).
It can be noticed that a weak but appreciable variation of $\tilde{\Phi}$
with $\phi$ is present.

\section{Summary and discussion}
\label{sec:discussion}

In the present work we have considered the problem of impurity particle transport 
in stellarator geometry. Impurity accumulation
can occur under certain plasma conditions and is predicted by the 
standard local and mono-energetic neoclassical theory.\\
Nevertheless, the mono-energetic approximation may 
not hold for impurities, whose high charge makes them sensitive 
to electrostatic potential variations within the flux surface.
This makes the problem at least -- assuming radial locality 
of the trajectories in the lowest order -- 
a 4D problem. The task of solving it to obtain the impurity particle flux density 
has been considered by the Monte-Carlo code \texttt{EUTERPE}.
The potential $\tilde{\Phi}(\theta)$ arising 
by enforcing quasi-neutrality in the code \texttt{GSRAKE} 
was used as input for the \texttt{EUTERPE} runs.\\
The results have shown that $\tilde{\Phi}$ can both 
increase and decrease the impurity accumulation,
depending on the interplay between 
the radial drives $\mathbf{v}_{\tilde{\Phi}}$ and $\mathbf{v}_{\mathrm{d}}$
and the electrostatic and magnetic trapping. These, in turn, 
have shown to be determined ultimately by the 
spectrum of $\tilde{\Phi}$ rather than by its absolute value. 
The calculations show that the inward flux of impurities 
can be suppressed completely.\\
Future work aims at a self-consistent calculation of $\tilde{\Phi}$
by \texttt{EUTERPE} along the line that
has been presented. The preparation of a global neoclassical version of the code, 
which would allow $E_{r}$ to be computed, is also forthcoming.

\section{Acknowledgements}
This work was supported by EURATOM and carried out within the framework of the European Fusion Development Agreement. The views and opinions expressed herein do not necessarily reflect those of the European Commision.\\
Part of the calculations were carried out using the HELIOS supercomputer system at Computational Simulation Centre of International Fusion Energy Research Centre (IFERC-CSC), Aomori, Japan, under the Broader Approach collaboration between Euratom and Japan, implemented by Fusion for Energy and JAEA. This work was also granted access to the HPC resources of HPC-FF made available within the Distributed European Computing Initiative by the PRACE-2IP, receiving funding from the European Community's Seventh Framework Programme (FP7/2007-2013) under grant agreement no. RI-283493.\\
J M Garc\'ia-Rega\~na wishes to thank K. Kauffmann for her helpful comments and the second referee for the constructive suggestions.

\section*{References}
\label{Bibliography}
\bibliographystyle{unsrt}

\end{document}